\documentclass{article}
\usepackage{authblk}
\usepackage[numbers]{natbib}
\usepackage{amsmath}
\usepackage{graphicx}
\usepackage{dcolumn}
\usepackage{bm}
\usepackage{rotating}
\usepackage[capitalize]{cleveref}
\usepackage{graphicx}
\usepackage{subfig}
\usepackage[utf8]{inputenc}
\usepackage[english]{babel}
\usepackage{color}
\usepackage{xspace}
\usepackage{soul}
\usepackage{ragged2e}
\usepackage{tikz}
\usepackage{tkz-tab}
\usepackage{pgfplots}
\pgfplotsset{compat=newest}

\begin{document}
\newlength\figureheight
	\newlength\figurewidth

\title{Hydrodynamic behavior of the Pseudo-Potential lattice Boltzmann method for interfacial flows}

\author[1]{Daniele Chiappini}
\author[2,4]{Xiao Xue}
\author[3]{Giacomo Falcucci}
\author[2]{Mauro Sbragaglia}
\affil[1]{University of Rome 'Niccolò Cusano', Department of Industrial Engineering, Via don Carlo Gnocchi 3, 00166, Rome, Italy, daniele.chiappini@unicusano.it}
\affil[2]{University of Rome 'Tor Vergata', Department of physics and INFN, Via della Ricerca Scientifica 1, 00133 Rome Italy}
\affil[3]{University of Rome 'Tor Vergata', Department of Enterprise Engineering “Mario Lucertini", Via del Politecnico 1 1, 00133 Rome Italy}
\affil[4]{Department of Physics, Eindhoven University of Technology, 5600 MB Eindhoven}

\maketitle

\begin{abstract}
The lattice Boltzmann method (LBM) is routinely employed in the simulation of complex multiphase flows comprising bulk phases separated by non-ideal interfaces. LBM is intrinsically mesoscale with an hydrodynamic equivalence popularly set by the Chapman-Enskog analysis, requiring that fields slowly vary in space and time. The latter assumptions become questionable close to interfaces, where the method is also known to be affected by spurious non hydrodynamical contributions. This calls for quantitative hydrodynamical checks. \textcolor{black}{In this paper we analyze} the hydrodynamic behaviour of \textcolor{black}{LBM pseudo-potential models for} the problem of break-up of a liquid ligament triggered by the Plateau-Rayleigh instability. Simulations are performed at fixed interface thickness, while increasing the ligament radius, i.e. in the "sharp interface" limit. Influence of different LBM collision operators is also assessed. We find that different distributions of spurious currents along the interface may change the outcome of the \textcolor{black}{pseudo-potential model} simulations quite sensibly, which suggests that a proper fine-tuning of \textcolor{black}{pseudo-potential models} in time-dependent problems is needed before the utilization in concrete applications. Taken all together, we argue that the results of the proposed study provide a valuable insight for engineering \textcolor{black}{pseudo-potential model} applications involving the hydrodynamics of liquid jets.
\end{abstract}

\section{INTRODUCTION}
The development of modern applications and innovative materials involving multiphase flows~\cite{yeoh2015handbook,atomization2017} naturally sets a compelling case for the development of suitably designed numerical methods to be used in synergy with experimental investigations~\cite{kuo1996} and analytical predictions~\cite{lin2003breakup}. The understanding of many of such problems is routinely rationalized via the help of a continuum hydrodynamics: in a nutshell one can say that bulk phases coexist while being separated by thin interfaces, whose width represents the smallest scale of the continuum description. Such interfaces are characterized by a non zero surface-tension, i.e. the force per unit area that is the continuum manifestation of the anisotropy of atomistic forces close to the interface. Whereas for purely analytical calculations the zero-width limit ("sharp interface" hydrodynamics) is most easily handable~\cite{eggers1993universal,eggers1994drop,magaletti2013sharp}, for numerical simulations the situation is somehow more diversified~\cite{eiken2006multiphase,ding2007diffuse,ding2007wetting,gibou2007level,yue2010sharp,worner2012numerical,luo2015conservative}. In this landscape, an increasing attention has been driven towards mesoscale simulations, and in particular the lattice Boltzmann method (LBM)~\cite{succi2018lattice,Qian,Higuera,Ansumali,Benzi,Aidun2010}. When solving the complex fluid dynamics of multiphase flows, the traditional advantages of the LBM (simplicity \cite{Falcucci2009,falcucci2017heterogeneous}, easy handling of boundary conditions~\cite{falcucci2016mapping,DiIlio2016b}, easy parallelization~\cite{chiappini2018numerical}) can be further enriched by a remarkable versatility in simulating non-ideal equation of states (EoS) and complex interfaces~\cite{Bella2009a,sbragaglia2007generalized}. More precisely, LBM reproduces "diffuse" interfaces, i.e. the bulk phases are separated by a region of finite thickness where the fluid properties (i.e. density, velocity, pressure) change continuously. The hydrodynamical behaviour of LBM is traditionally assessed via the Chapman-Enskog analysis; however, from the theoretical point of view, the main assumptions of the Chapman-Enskog analyis of having fields slowly varying in space and time may well be violated due to the presence of the interfaces and/or singular events like break-ups~\cite{shan1995multicomponent, zheng2006lattice}. Practically, it is also found that LBM implementations are affected by spurious contributions at the interface~\cite{sbragaglia2009lattice,Colosqui2012}. We use the term "spurious" meaning that they are not predicted by hydrodynamics. These spurious currents are particularly relevant close to the interfaces~\cite{shan2006analysis}. Consequently, the actual recovery of the LBM-hydrodynamic "equivalence" could fail~\cite{succi2018lattice,kruger2017lattice}. Natural questions then arise, on the quantitative potentiality retained by diffuse interface LBM simulations of multiphase flows, especially in comparison to the analytical description of sharp interface hydrodynamics. In fact, while it is largely acknowledged in the literature~\cite{:ShanChen,Lee2006a,Falcucci,Sbragaglia2006a, Premnath2007,Kupershtokh2009,Aidun2010,Falcucci2010,li2013lattice,Zhang2014,ammar2017multiphase} that LBM is capable of reproducing static properties driven by surface tensions (i.e. Laplace pressures~\cite{sbragaglia2013interaction}, contact angles~\cite{Sbragaglia2006}), very rarely there have been quantitative characterizations on the recovery of time-dependent hydrodynamics with non-ideal interfaces, especially in presence of singular events. This paper aims to take a step forward in the latter direction. As a prototypical problem of a time-dependent multiphase flow with non-ideal interfaces we refer to the Plateau-Rayleigh instability~\cite{Plateau,rayleigh1878instability,rayleigh1882further,chandrasekhar1961hydrodynamic,rutland1970theoretical,tjahjadi1992satellite,eggers1994drop,ashgriz1995temporal} of a liquid ligament. The Plateau-Rayleigh instability -- driven by the tendency of the interface to minimize the area at fixed available volume -- causes the fragmentation of a liquid ligament into smaller droplets via break-up events. Numerical results on the break-up time show a neat asymptotic behaviour when the interface width is much smaller than the ligament size. These asymptotic results are compared with the theoretical predictions of sharp interface hydrodynamics; moreover, our observations are also enriched with a side-by-side comparison of two different LBM collision operators, namely the Single Relaxation Time (SRT) with shifted equilibrium~\cite{sbragaglia2009lattice} (hereafter SRT), and Multiple Relaxation Time (MRT) with Guo-like forcing~\cite{guo2002discrete} (hereafter MRT). Numerical simulations show that the distribution of velocity at the interface in the vicinity of the pinch-off region is different, causing different break-up processes. A very preliminar investigation on some of the results presented in this paper is also available in a recent conference proceedings~\cite{chiappini2018ligament}.\\
The paper is organized as follows. In Section~\ref{sec:LBM} we recall the basic features of the numerical methodology used; in Section~\ref{sec:setup} we report on the set-up used for the numerical simulations; results will be presented in Section~\ref{sec:results} and in Section~\ref{sec:conclusions} conclusions will be drawn.
\section{NUMERICAL MODELS}\label{sec:LBM}
In this section, we briefly highlight the important features of the numerical methodology based on the \textcolor{black}{LBM}~\cite{Benzi,succi2018lattice}. For extensive technical details the interested reader is referred to the papers cited in the following. LBM is a mesoscale numerical approach for the study of fluid dynamics, which has been successfully employed to dissect complex phenomena of scientific and technical interest in recent years~\cite{sbragaglia2007generalized,Aidun2010,kruger2011efficient,Sbragaglia2006a,Falcucci2010,falcucci2013direct,falcucci2016mapping}. \\
LBM is grounded on an optimized formulation of Boltzmann's kinetic equation, in which particle distribution functions $f_\alpha\left(\textbf{x},t\right)$ stream and collide on a lattice characterized by a finite set of velocities $\textbf{c}_{\alpha} = 0,...,18$ in our case, according to the following dynamics:
\begin{equation}\label{eq:LBE}
\Delta_\alpha f_\alpha\left(\textbf{x},t\right) = f_\alpha\left(\textbf{x}+\textbf{c}_\alpha\Delta t,t+\Delta t\right)-f_\alpha\left(\textbf{x},t\right) = -\Omega_{\mbox{\tiny coll}} \left[f_\alpha\left(\textbf{x},t\right) - f_\alpha^{\mbox{\tiny eq}}\left(\textbf{x},t\right)\right].
\end{equation}
In Eq.~\ref{eq:LBE} $\Omega_{\mbox{\tiny coll}}$ represents the collision operator, which can be written as follows:
\[
\left\{
\begin{array}{l l}
\Omega_{\mbox{\tiny coll}} &= \Delta t/\tau \quad \text{(\mbox{Single Relaxation Time, SRT})};  \\
\Omega_{\mbox{\tiny coll}} &= \bm{M \ \Lambda \ M^{-1}} \quad \text{(\mbox{Multiple Relaxation Time, MRT})}:   \\
\end{array}
\right.
\label{collis_op}
 \]
$\tau$ represents the (single) relaxation time towards local equilibrium~\cite{Benzi}; $\bm{M}$ and $\bm{\Lambda}$ are the transformation matrix and the (diagonal) matrix of the relaxation parameters, respectively: for the details on thier formulation, the reader is addressed to~\cite{ammar2017multiphase}. More specifically, the $\bm{\Lambda}$ vector has $s_2=s_{10}=s_{12}=s_{14}=s_{15}=s_{16}=\omega=\frac{2c_s^2}{c_s^2 + \nu}$ and all the other free parameters equal to 1, where $\nu$ is the kinematic viscosity and $c_s^2$ is the squared lattice speed of sound. No special adjustments have been considered here, because modifications on $\bm{\Lambda}$ vector should not affect hydrodynamic behaviour of the employed method. For both the SRT and MRT formulations, the term $f_\alpha^{\mbox{\tiny eq}}\left(\textbf{x},t\right)$ in Eq.~\ref{eq:LBE} represents the distribution of (local) Maxwellian equilibrium, which is given by
\begin{equation}
\label{eq:local_eq}
f_\alpha^{\mbox{\tiny eq}}\left(\textbf{x},t\right) =  w_\alpha\rho\left(\textbf{x},t\right) \bigg[\frac{\textbf{c}_\alpha\cdot\textbf{u}\left(\textbf{x},t\right)}{c_s^2}+\frac{\left[\textbf{c}_\alpha\cdot\textbf{u}\left(\textbf{x},t\right)\right]^2}{2c_s^4}  - \frac{\left[\textbf{u}\left(\textbf{x},t\right)\cdot\textbf{u}\left(\textbf{x},t\right)\right]}{2c_s^2}\bigg] \; ,
\end{equation}
in which $\rho\left(\textbf{x},t\right)$ and $\textbf{u}\left(\textbf{x},t\right)$ are the hydrodynamic macroscopic density and velocity, respectively and $w_{\alpha}$ represents the set of weights for the D3Q19 lattice~\cite{succi2018lattice,kruger2017lattice}. From Eq.~\ref{eq:LBE}, macroscopic fluid density and velocity may be derived through the $0^{th}$ and the $1^{st}$ population momentum, respectively, as follows:
\begin{equation}
	\label{eq:momentum}
	\rho\left(\textbf{x},t\right)=\sum_{\alpha=0}^{N_{\mbox{\tiny pop}}-1}f_\alpha\left(\textbf{x},t\right) \;\;\;\;\;\;\;\;\;\;\;	\rho\textbf{u}\left(\textbf{x},t\right)=\sum_{\alpha=0}^{N_{\mbox{\tiny pop}}-1}\textbf{c}_\alpha f_\alpha\left(\textbf{x},t\right).
\end{equation}
One among the main interesting \textit{atouts} of the LBM lays in its effectiveness in dealing with non-ideal, multiphase flows~\cite{Aidun2010}: the forcing term can be conveniently implemented in Eq.~\ref{eq:LBE} to account for the phase interactions that trigger the macroscopic phase segregation. Among the various approaches proposed in the literature, we focus on the single-belt formulation of the \textcolor{black}{pseudo-potential} Shan-Chen forcing~\cite{:ShanChen,shan1994simulation}, whose force reads
\begin{equation}\label{SC_force}
\textbf{F}\left(\textbf{x},t\right) = -{G_0}\psi \left(\textbf{x},t\right)\sum_{\alpha=0}^{N_{\mbox{\tiny pop}}-1}\psi \left(\textbf{x}+\textbf{c}_{\alpha}\Delta t,t\right)\textbf{c}_{\alpha}w_{\alpha} \; .
\end{equation}
In Eq.~\ref{SC_force}, $G_0$ is the basic parameter which rules the inter-particle interaction and $\psi\left(\textbf{x},t\right)$ is the \textit{pseudo-potential}, a local functional of the fluid density~\cite{chen1998lattice}:
$$
\psi\left(\textbf{x},t\right) = \rho_0\left[1 - \exp\left(-{\rho\left(\textbf{x},t\right)\over{\rho_0}}\right)\right].
$$
In this work, we have fixed the reference density $\rho_0 = 1.0$: with this assumption, the inter-particle strength $G_0$ is the only free parameter which fixes both the density ratio and the surface tension. Forcing schemes are different for the two collision operators; more in detail, for the SRT, starting from equation~\ref{SC_force}, the component of the interaction potential along each direction can be evaluated and then used to shift the macroscopic velocities before evaluating the equilibrium distribution functions:
\begin{equation}\label{vel_shift}
\textbf{u}^{'}\left(\textbf{x},t\right) = \textbf{u}\left(\textbf{x},t\right) + {{\textbf{F}\left(\textbf{x},t\right)\tau}\over{\rho\left(\textbf{x},t\right)}}.
\end{equation}
On the other hand, for the MRT scheme, we adopted the forcing scheme proposed in~\cite{Premnath2007, Zhang2014} where we compute the equilibrium momentum by means of the velocity evaluated as follows:
\begin{equation}
\textbf{u}^{b}\left(\textbf{x},t\right) = \textbf{u}\left(\textbf{x},t\right) + {{\textbf{F}\left(\textbf{x},t\right)}\over{2\cdot\rho\left(\textbf{x},t\right)}}.
\label{eq:barycentric}
\end{equation}
For both the considered collision operators, the EoS of the system may be written as follows:
\begin{equation}\label{EOS_SC}
P(\rho) = \rho c_s^2 + {{c_s^2G_0}\over{2}}\psi^2.
\end{equation}
Before ending this methodological section, some remarks on the models used are in order. The main difference in between the two collision operators is that SRT is solved into space, while MRT projects the distribution functions into momentum space, by means of the matrix $\textbf{M}$ product. This technical passage allows to increase the stability and robustness of the method itself. Regarding the equilibrium properties (i.e. density ratio, surface tension, etc etc), it is well known that the basic Shan-Chen formulation is affected by some pathologies. Indeed, in the SRT formulation, both the surface tension and density ratio depend on the relaxation parameter $\tau$ \cite{shan1994simulation}, while in the MRT formulation they are decoupled from it. This is confirmed by Tables~\ref{tab:radius} and~\ref{tab:radius_MRT}. Furthermore, due to the forcing formulation, the surface tension is a function of the parameter $G_0$ itself, which causes a coupling between the EoS and the interface properties and results in the impossibility to tune surface tension independently of the EoS. Some extensions of the basic Shan-Chen force were designed to cure such pathology. Sbragaglia {\it et al.} \cite{sbragaglia2007generalized} and Falcucci {\it et al.} \cite{Sbragaglia2006a} proposed to extend the range of interactions (i.e. multirange approach) of the Shan-Chen forces, allowing an independent tuning of the surface tension with respect to the EoS. Other studies followed, aimed at systematic characterizations and further improvements. For example, Yu \& Fan \cite{yu2009interaction} used a multirange approach to allow for non uniform meshes and grid-refinement close to non-ideal interfaces; Huang {\it et al.} \cite{huang2011forcing} systematically analyzed the impact of the multirange formulation on the equilibrium properties of non-ideal interfaces; Li {\it et al.} \cite{li2013achieving} proposed a modified approach by adding a source term to the LBM allowing the independent tuning of surface tension with respect to the EoS. The multirange extensions are here not explored; rather, we focus on the basic formulation of the pseudo-potential approach to delve some general considerations about "dynamical" spurious currents effects on macroscopic hydrodynamic phenomena. Indeed, the two LBM environments are here chosen as "representative" of two scenarios, where spurious hydrodynamical effects exhibit a different modulation in space, in one case more localized in the pinching region, in another case localized away from the pinching region. In other words, the two LBM environments are not chosen to promote one with respect to the other, but rather to raise a more general question on the dynamical distribution of non-hydrodynamical effects.

%
\section{SIMULATIONS SET-UP}\label{sec:setup}
\begin{figure}
\begin{center}
\includegraphics[width=0.8\columnwidth]{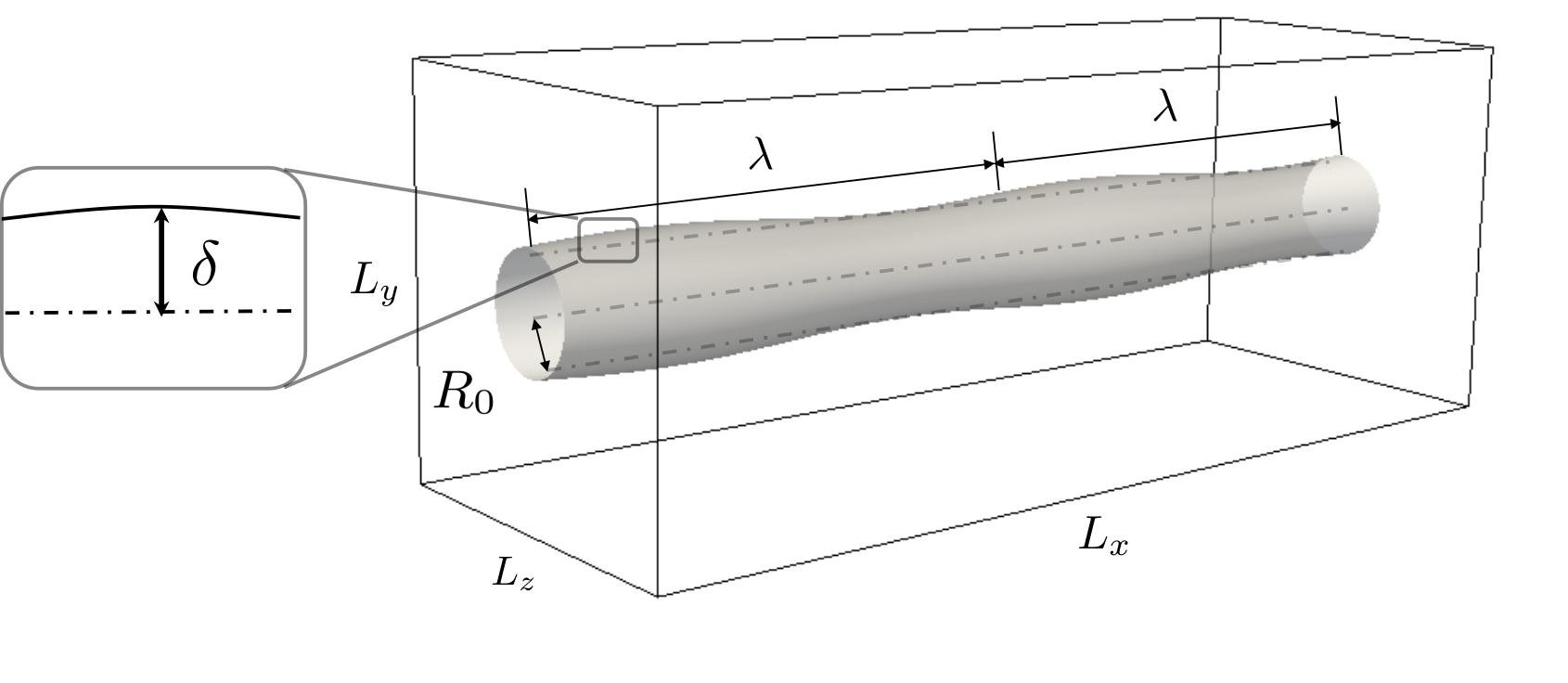}
\end{center}
\caption{Initial configuration for the numerical simulations. A cylindrical ligament of radius $R_0$ is perturbed with a sinusoidal wave along the axial ($x$) coordinate. The perturbation wavelength corresponds to the fastest growing mode of the Plateau-Rayleigh instability~\cite{eggers1994drop}.}
\label{fig:ligament_init}
\end{figure}

%
We have performed LBM simulations in a 3D box of $L_x \times L_y \times L_z$ lattice sites. The Plateau-Rayleigh instability is triggered through a sinusoidal perturbation with a fixed amplitude and a constant wavelength (see Fig.~\ref{fig:ligament_init}). More specifically, the initial condition for the rigament radius is $r\left(x\right) = R_0 + \delta \sin\left(\frac{4 \pi x}{L_x}\right)$, where $R_0$ is the unperturbed ligament radius. The domain size and the ligament radius are chosen to accommodate roughly 2 wavelengths of the most unstable mode of the instability~\cite{eggers1994drop}. The perturbation $\delta$ is assigned the three different values $R_0/30$, $R_0/20$, $R_0/10$. The break-up phenomenon is driven by some characteristic parameters, between them the Ohnesorge number $\mbox{Oh}$ and the capillary time $t_{\mbox{\tiny cap}}$ defined as follows:
\begin{equation}
\mbox{Oh} = \frac{\mu}{\sqrt{\rho_l\sigma R_0}} \hspace{.2in} t_{\mbox{\tiny cap}} = \sqrt{\frac{\rho_l R_0^3}{\sigma}}
\label{eq:Oh_Tcap}
\end{equation}
where $\rho_l$ is the liquid density, $\mu$ the dynamic viscosity, $\sigma$ the surface tension. For our numerical simulations, the inter-particle strength $G_0$~\cite{sbragaglia2007generalized,Sbragaglia2006a} has been fixed to $-5.3$ LU (lattice units hereafter), which allows a fair grid convergence study at changing the ligament radius $R_0$, from values comparable to the interface thickness to values much larger. Regarding the choice of the Ohnesorge number, a few remarks are in order. According to the literature on the break-up of liquid ligaments~\cite{schulkes1996contraction,eggers2008physics,driessen2013stability}, it is known that for $\mbox{Oh}<1$ instability phenomena of the liquid jet start to take place, causing the formation of "pinched" regions, which, eventually, lead to the break-up of the liquid column. To accomplish our analysis in the LBM framework, we have fixed $\mbox{Oh} = 0.1$, which is a well-established value on the "pinching" break-up regime~\cite{notz2004dynamics}. Moreover, through such a value, it is possible to have a set of corresponding numerical viscosities that grant the numerical stability of the LBM algorithm~\cite{succi2018lattice,kruger2017lattice}, as explained in the following. The ligament radius $R_0$ has been varied in the range $14-98$ LU. Table~\ref{tab:radius} reports the corresponding values of density ratio and surface tension retrieved with the SRT approach, for the different values of $R_0$. To accurately evaluate the surface tension $\sigma$ as a function of the ligament radius, we first carried out a set of steady-state simulations to perform the well known Laplace test, \cite{de2013capillarity}. With the SRT collision operator, Laplace test requires to account for the natural adaptation of $\sigma$ to the kinematic viscosity value~\cite{shan1994simulation}, while in the MRT framework such an effect is absent. To retrieve a reliable value for the surface tension for all the ligament radii reported in Table~\ref{tab:radius}, we have implemented an iterative procedure aimed at providing stable values for $\sigma$. More specifically, we chose a first attempt estimation for the viscosity, aimed at ensuring numerical stability; with the corresponding $\tau$, we have performed the Laplace test according to three values of the static droplet radius, obtaining a first estimate of $\sigma$. By using the obtained value of surface tension in Eq.~\ref{eq:Oh_Tcap}, we find the new viscosity value that meets the $\mbox{Oh}=0.1$ target: since the values of $\sigma$, $\rho_l$ and $\rho_v$ (vapour density) are intimately connected to that of $\tau$ in the SRT approach, we have iterated the above-described procedure until the relative error on two consecutive values of surface tension was less than 5$\%$. For the MRT, no such procedure was needed, as viscosity variations have a negligible impact on the algorithm. Once we acquired the asymptotic surface tension value for both collision operators, we performed simulations with flat interfaces to find the correct liquid and vapour densities corresponding to the simulation parameters. Finally, after this last set of simulations we obtained all the input data needed to perform the ligament simulations. Table~\ref{tab:radius_MRT} displays the values of density ratio and surface tension obtained with the MRT collision operator: as it is apparent \textcolor{black}{from} the comparison between Tables~\ref{tab:radius} and~\ref{tab:radius_MRT}, the values provided by the MRT display a negligible dependence of the physical properties on the employed computational grid.\\ 
Before closing this section, we notice that the break-up of a thin ligament has already been studied by means of a axisymmetric LBM formulation in~\cite{PremnathAbraham2005,srivastava2013, srivastava2013lattice,Liangetal14,Liuetal16}; in this paper, we extend the results already available in the literature by performing a 3D grid convergence analysis, by analysing the effect of different collision operators on the predictions of hydrodynamics and by performing a more detailed analysis on the properties of the break-up process.\\
\begin{table*}[t!]
\caption{Main simulation parameters at $G_0=-5.3$ and $\mbox{Oh}=0.1$ as a function of ligament radius for SRT.}
\label{tab:radius}
\tabcolsep7pt\begin{tabular}{l c c c c c c c}
\hline
 $R_0$ (LU) & 14 & 28 & 42 & 56 & 70 & 84  & 98\\
 \hline
$L_x$ (LU) & 256 & 512 & 768 & 1024 & 1280 & 1536 & 1792\\
$L_y=L_z$ (LU) & 96 & 192 & 288 & 384 & 480 & 576 & 672\\
$\tau$ (LU)& 0.67926 & 0.76297 & 0.83053 & 0.90311 & 0.97021 & 1.03776 & 1.10647\\
$t_{\mbox{\tiny cap}}$ (LU) & 335 & 912 & 1611 & 2380 & 3189 & 4015 & 4846\\
$\rho_l$ (LU) & 2.1167 & 2.1254 & 2.1348 & 2.1450 & 2.1557 & 2.1666 & 2.1775\\
$\rho_v$ (LU) & 0.0836 & 0.0900 & 0.0974 & 0.1059 & 0.1156 & 0.1264 & 0.1383\\
$\sigma$ (LU) & 0.05345 & 0.05775 & 0.06277 & 0.06848 & 0.07491 & 0.08206 & 0.08991\\
\hline
\end{tabular}
\end{table*}
\begin{table*}[t!]
\caption{Main simulation parameters at $G_0=-5.3$ and $\mbox{Oh}=0.1$ as a function of ligament radius for MRT.}
\label{tab:radius_MRT}
\tabcolsep7pt\begin{tabular}{l c c c c c c c}
\hline
 $R_0$ (LU)  & 14 & 28 & 42 & 56 & 70 & 84  & 98\\
 \hline
$L_x$ (LU)  & 256 & 512 & 768 & 1024 & 1280 & 1536 & 1792\\
$L_y=L_z$ (LU) & 96 & 192 & 288 & 384 & 480 & 576 & 672\\
$\tau$ (LU) & 0.67293 & 0.74445 & 0.79942 & 0.84575 & 0.88667 & 0.92350 & 0.95745\\
$t_{\mbox{\tiny cap}}$ (LU) & 347 & 981 & 1803 & 2775 & 3878 & 5098 & 6424\\
$\rho_l$ (LU) & 2.1086 & 2.1086 & 2.1086 & 2.1086 & 2.1086 & 2.1085 & 2.1085\\
$\rho_v$ (LU) & 0.0778 & 0.0779 & 0.0779 & 0.0779 & 0.0778 & 0.0778 & 0.0778\\
$\sigma$ (LU) & 0.04951 & 0.04952 & 0.04952 & 0.04955 & 0.04953 & 0.04953 & 0.04954\\
\hline
\end{tabular}
\end{table*}
%
\section{RESULTS}\label{sec:results}
\begin{figure}[t!]
\begin{minipage}{0.33\columnwidth}
\begin{center}
(a)\\
\captionsetup[subfigure]{labelformat=empty}
\subfloat[\tiny $R_0 = 28 \left(\mathrm{LU}\right) ; t^\star = 7$]{\includegraphics[width = 0.45\columnwidth]{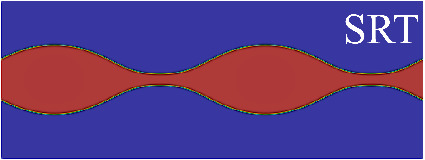}}
\captionsetup[subfigure]{labelformat=empty}
\subfloat[\tiny $R_0 = 98 \left(\mathrm{LU}\right) ; t^\star = 7$]{\includegraphics[width = 0.45\columnwidth]{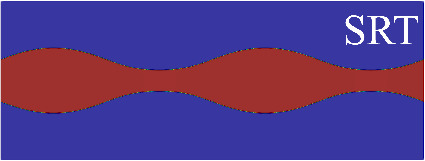}}\\
\captionsetup[subfigure]{labelformat=empty}
\subfloat[\tiny $R_0 = 28 \left(\mathrm{LU}\right) ; t^\star = 8$]{\includegraphics[width = 0.45\columnwidth]{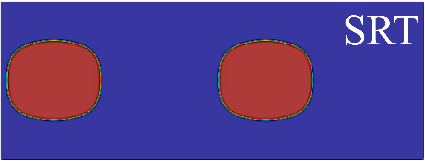}}
\captionsetup[subfigure]{labelformat=empty}
\subfloat[\tiny $R_0 = 98 \left(\mathrm{LU}\right) ; t^\star = 8$]{\includegraphics[width = 0.45\columnwidth]{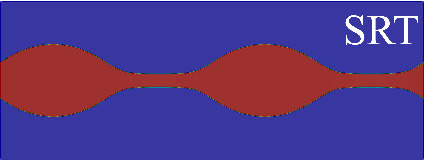}}\\
\captionsetup[subfigure]{labelformat=empty}
\subfloat[\tiny $R_0 = 28 \left(\mathrm{LU}\right) ; t^\star = 10$]{\includegraphics[width = 0.45\columnwidth]{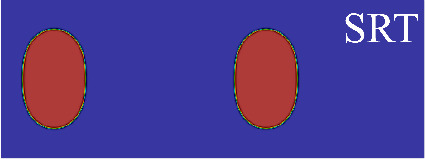}}
\captionsetup[subfigure]{labelformat=empty}
\subfloat[\tiny $R_0 = 98 \left(\mathrm{LU}\right) ; t^\star = 10$]{\includegraphics[width = 0.45\columnwidth]{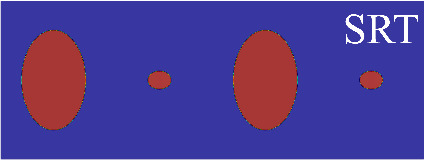}}
\end{center}
\end{minipage}
\begin{minipage}{0.35\columnwidth}
\begin{center}
(b)\\
\vspace{0.1cm}
\setlength\figureheight{0.7\columnwidth}
\setlength\figurewidth{0.88\columnwidth}
%
%
\begin{tikzpicture}
\begin{axis}[%
width=\figurewidth,
height=\figureheight,
scale only axis,
xmin=10,
xmax=100,
xtick={10,20,30,40,50,60,70,80,90,100},
xlabel={$R_0$ [LU]},
ymin=5,
ymax=10,
ytick={5,6,7,8,9,10},
ylabel={$T_{\mathrm{\tiny break}}$ [-]},
legend style={at={(1,0)}, anchor=south east,legend cell align=left}
]
\addplot [color=blue,line width=1.0pt,solid,mark size=4pt,only marks,mark=square,mark options={solid}]
  table[row sep=crcr]{%
14	5.975\\ 
28	7.375\\
42	7.875\\
56	8.250\\
};	
\addlegendentry{SRT};
\addplot [color=red,line width=1.0pt,solid,mark size=4pt,only marks,mark=triangle,mark options={solid}]
  table[row sep=crcr]{%
14	5.925\\    
28	6.925\\
42	7.425\\
56	7.850\\
70	8.125\\
84	8.225\\
98	8.350\\
};
\addlegendentry{MRT};
\addplot [color=blue,line width=1.0pt,solid,mark size=4pt,only marks,mark=square*,mark options={fill=blue}]
  table[row sep=crcr]{%
70	8.525\\
84	8.675\\
98	8.825\\
};
\end{axis}
\end{tikzpicture}%
\end{center}
\end{minipage}\hspace{0.1cm}
\begin{minipage}{0.27\columnwidth}
\begin{center}
(c)\\
\vspace{0.4cm}
\captionsetup[subfigure]{labelformat=empty}
\subfloat[\tiny SRT - $t^\star = 10$]{\includegraphics[width = 0.75\columnwidth]{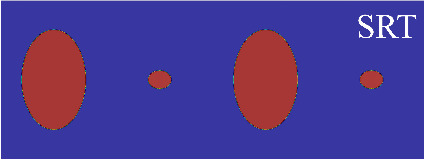}}\\
\vspace{0.5cm}
\captionsetup[subfigure]{labelformat=empty}
\subfloat[\tiny MRT - $t^\star = 10$]{\includegraphics[width = 0.75\columnwidth]{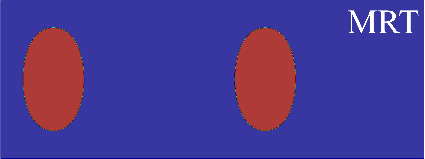}}
\vspace{0.3cm}
\end{center}
\end{minipage}
\caption{Panel (a): we report the LBM density evolution as a function of the non-dimensional time ($t^{\star}$) for the SRT collision operator and for different resolutions. Panel (b): we report the dimensionless break-up times as a function of the ligament radius for both SRT and MRT. Filled symbols refer to the presence of satellite droplets in post break-up conditions. Panel (c): post break-up conditions for both SRT and MRT. \label{fig:bu_time}}
\end{figure}
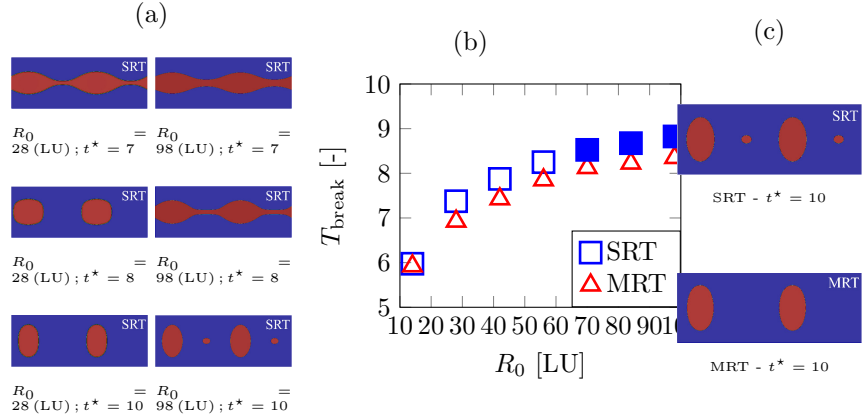

\begin{figure}[h!]
\begin{center}
	\setlength\figureheight{0.4\columnwidth}
	\setlength\figurewidth{0.6\columnwidth}
%
%
\begin{tikzpicture}
\begin{axis}[%
width=\figurewidth,
height=\figureheight,
scale only axis,
log ticks with fixed point,
xmin=0.002,
xmax=0.122,
xtick={0.002, 0.022, 0.042, 0.062, 0.082, 0.102, 0.122},
xlabel={$\epsilon = \frac{\delta}{R_0}$ [-]},
xticklabel style={/pgf/number format/.cd,fixed,precision=3},
ymin=6,
ymax=24,
ytick={6, 9, 12, 15, 18, 21},
ylabel={$T_{\mathrm{break}}$ [-]},
legend style={at={(1,1)}, anchor=north east,legend cell align=left}
]
\addplot [color=black,line width=2.0pt,solid,mark size=4pt,only marks,mark=diamond*,mark options={solid}]
  table[row sep=crcr]{%
0.05	11.70124944	\\
};
\addlegendentry{Ashgriz et. al [52]};
\addplot [color=purple,line width=2.0pt,dashed]
  table[row sep=crcr]{%
0.002	22.19502892	\\
0.007	17.72087546	\\
0.012	15.79588796	\\
0.017	14.55193548	\\
0.022	13.63111723	\\
0.027	12.89970862	\\
0.032	12.29292634	\\
0.037	11.77441917	\\
0.042	11.3217345	\\
0.047	10.92002742	\\
0.052	10.55896986	\\
0.057	10.23108575	\\
0.062	9.930788907	\\
0.067	9.653795213	\\
0.072	9.396747	\\
0.077	9.156963775	\\
0.082	8.932271542	\\
0.087	8.720882715	\\
0.092	8.52130965	\\
0.097	8.332301073	\\
0.102	8.15279452	\\
0.107	7.981880159	\\
0.112	7.818772885	\\
0.117	7.662790515	\\
0.122	7.513336551	\\
};
\addlegendentry{Linear Stability Analysis [49]};
\addplot [color=blue,line width=2.0pt,solid,mark size=4pt,only marks,mark=square,mark options={solid}]
  table[row sep=crcr]{%
0.033333333	13.3	\\
0.05	11.65	\\
0.1	8.825	\\
};
\addlegendentry{SRT};
\addplot [color=red,line width=2.0pt,solid,solid,mark size=4pt,only marks,mark=triangle,mark options={solid}]
  table[row sep=crcr]{%
0.033333333	12.175	\\
0.05	10.8	\\
0.1	8.35	\\
};
\addlegendentry{MRT};

\end{axis}
\end{tikzpicture}%
	\end{center}
	\caption{Influence of the initial perturbation $\epsilon = \frac{\delta}{R_0}$ for the two collision operators on the non-dimensional break-up time and comparison with literature data~\cite{ashgriz1995temporal} and linear stability analysis prediction~\cite{chandrasekhar1961hydrodynamic}.\label{HYDROCOMPARISON}}
\end{figure}
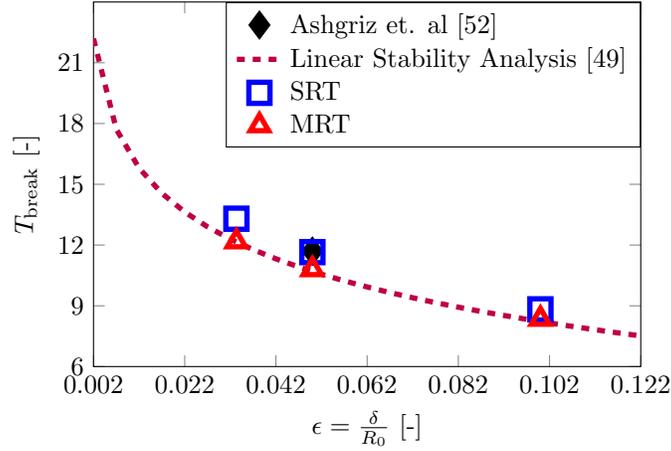

\begin{figure}[h!]
\begin{minipage}{0.5\columnwidth}
\begin{center}
(a)\\
\includegraphics[height=0.94\columnwidth]{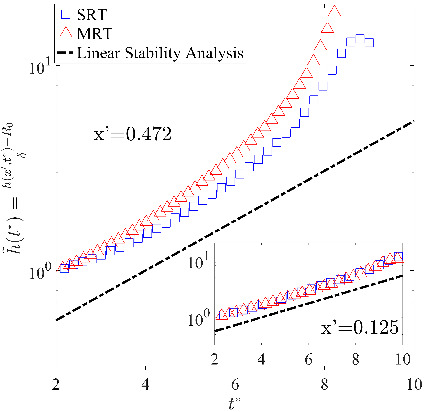}
\end{center}
\end{minipage}
\begin{minipage}{0.40\columnwidth}
	\begin{center}
	(b)\\
	\setlength\figureheight{0.30\columnwidth}
	\setlength\figurewidth{0.85\columnwidth}
	\input{./int_t7.txt}\\
	\setlength\figureheight{0.30\columnwidth}
	\setlength\figurewidth{0.85\columnwidth}
	\input{./int_t8.txt}\\
	\setlength\figureheight{0.30\columnwidth}
	\setlength\figurewidth{0.85\columnwidth}
	\input{./int_t9.txt}\\
	\end{center}
\end{minipage}
\caption{\label{profiles} Panel (a): Normalized perturbation with respect to initial perturbation value at $x \approx L_x/2$ (and $x = L_x/8$ in the inset) as a function of non-dimensional time. Panel (b): Three snapshots of interface evolution at $t^\star=7,8,9$ both for SRT and MRT.}
\end{figure}

\begin{figure}[h!]
\begin{center}
	\setlength\figureheight{0.4\columnwidth}
	\setlength\figurewidth{0.5\columnwidth}
	\input{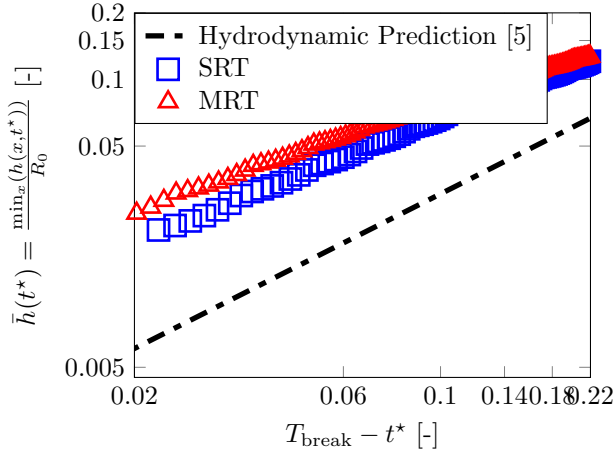}
	\end{center}
	\caption{Normalized minimum ligament height near the break-up time. \label{fig:linear_pinch}}
\end{figure}

\begin{figure}
\begin{center}
	\subfloat[]{\includegraphics[width=0.5\textwidth]{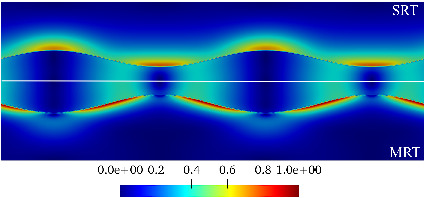}} \\
	\subfloat[]{\includegraphics[width=0.5\textwidth]{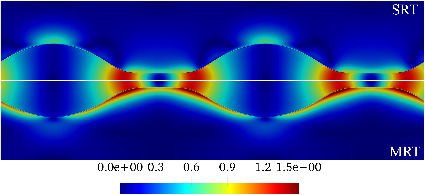}} \\
	\subfloat[]{\includegraphics[width=0.5\textwidth]{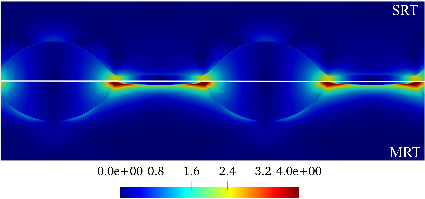}}
	\end{center}
	\caption{Comparison of the time evolution of the normalized velocity magnitude according to the SRT and MRT collision operators. \label{vel}}
\end{figure}

According to the parameters reported in Tables~\ref{tab:radius} and~\ref{tab:radius_MRT}, we have performed the numerical simulations for the break-up of the liquid ligament for $\delta=R_0/10$. In Fig.~\ref{fig:bu_time} we report results on the time evolution (up to the break-up point) and the break-up times $T_{\mbox{\tiny break}}$ at changing the grid resolution. Notice that the break-up time has been made dimensionless with respect to the capillary time $t_{\mbox{\tiny cap}}$. For both the collision operators, we observe a very neat trend of the break-time increasing with the simulation resolution and, thus, with the initial radius $R_0$, in line with the numerical results in~\cite{tiwari2008simulations}. In our working conditions, sharp interface hydrodynamics predicts a dimensionless break-up time that is a function only of $\mbox{Oh}$ and $\delta$ (see~\cite{driessen2011regularised} and references therein). Since $\mbox{Oh}$ and $\delta$ are fixed, the observed grid dependency cannot be explained in terms of sharp interface hydrodynamics and it is an effect induced by the finite width of the interface. It naturally comes the question of how much these finite width effects are "hydrodynamical". To cope with this issue, one would need to use a "diffuse" interface \textcolor{black}{hydrodynamic} solver for the corresponding hydrodynamic equations predicted by LBM~\cite{succi2018lattice,kruger2017lattice}. Finite width effects are expected to be negligible for large resolutions and one can use the results of sharp interface hydrodynamics (see~\cite{driessen2011regularised} and references therein) for comparison. For the Ohnnesorge number used, $\mbox{Oh}=0.1$, the break-up process is known to produce 2 mother droplets and 2 satellite droplets. In Fig.~\ref{fig:bu_time}(b), the cases in which \textit{stable} secondary  droplets are found are reported with filled symbols. We would like to stress that only \textit{stable} droplets have been considered, that is, secondary droplets that live for a period of time at least comparable to the capillary time $t_{\mbox{\tiny cap}}$. It is known, in fact, that in the pseudo-potential framework of LBM multi-phase flows, \textit{small} droplets tend to be re-absorbed in the vapor phase, as discussed in~\cite{chibbaro2008lattice,Falcucci2010}. In our simulations, SRT provides stable secondary droplets for $R_0 \geq 70$ LU, while the MRT collision operator provides secondary droplets that live considerably less than a single capillary time. Such a short-living feature is due to their initial diameter, which tends to be smaller with MRT than with SRT. To the best of the authors' knowledge, no previous works both with SRT and MRT, point out such a persistence characteristic of the secondary droplets~\cite{PremnathAbraham2005,srivastava2013}. By comparing the diameters obtained with the two collision operators to the analytic, numerical and experimental results in the literature~\cite{rutland1970theoretical,lafrance1975nonlinear,mansour1990satellite}, we find that the SRT approach provides more accurate predictions. This surely calls for a careful comparison with the existing LBM data in the literature. Earlier investigation by Premnath \& Abraham~\cite{PremnathAbraham2005} used an axisymmetric LBM formulation with source terms embedded in the LBM dynamics in the BGK approximation. The Carnahan-Starling EoS was used with a dynamic viscosity ratio of 4 between the liquid and vapor phase. The authors report the formation of satellite droplets with an initial cylinder radius of 50 grid points (their Fig. 7), hence well below our largest resolutions used.  Srivastava {\it et al.}~\cite{srivastava2013} proposed an axisymmetric LBM formulation with the Shan-Chen force and the forcing scheme provided by the equilibrium shift, essentially the same of our forcing scheme for the SRT simulations. For $\mbox{Oh}=0.09$ they reported the formation of satellite droplets with dynamic viscosity ratio of about 30, which is consistent with our results. Another axysimmetric LBM formulation proposed by Liang {\it et al.}~\cite{Liangetal14} makes use of a phase field Van der Waals model and adds the forcing term as a source to the BGK evolution. They use an initial cylinder radius with 60 grid points and report presence of satellite droplets with dynamic viscosity ratio of about 12.5. In the recent investigation by Liu {\it et al.}~\cite{Liuetal16}, the authors report numerical simulations with an axisymmetric LBM with a color-gradient model and MRT collision operator. Finally, in the recent simulations by some of the authors with the Shan-Chen force for a multicomponent fluid and MRT forcing scheme, it was found the emergence of the satellite droplets~\cite{xue2018fluctliga}. These facts said, it is likely that the missed formation of satellite droplets in our MRT simulations originates from the chosen EoS and the choice of a viscosity ratio that sensibly differs from one. \\
Although at a very qualitative level, results in Fig.~\ref{fig:bu_time} provide a clue to a physically different behaviour of the two collision operators, with SRT looking "more physical" than MRT. This result is strange and counter-intuitive, since SRT is known to be affected by extra forcing-dependent stress contributions, which are large at large forces (i.e. close to interfaces).
These observations stimulated further analysis on the quantitative comparisons between the numerical simulations and the theoretical predictions. To this aim we kept the resolution fixed to the largest used in Fig.~\ref{fig:bu_time} and investigated the break-up time at changing $\delta$. In Fig.~\ref{HYDROCOMPARISON} results of the numerical simulations are compared with the predictions of sharp interface hydrodynamics from Ref.~\cite{ashgriz1995temporal} and linear stability analysis. Regarding the latter, the growth rate $\omega$ is considered as a function of both the wavenumber and the Ohnesorge number (see Eq. (28) in~\cite{eggers1994drop}). Knowing the growth rate, the break-up time in the linear approximation can be calculated from $T_{\mbox{\tiny break}}=\log (R_0/\delta)/\omega$~\cite{chandrasekhar1961hydrodynamic,ashgriz1995temporal}. We notice that the MRT collision operator is well aligned with linear stability analysis reported in~\cite{chandrasekhar1961hydrodynamic}, while SRT is practically overlapped with result presented in~\cite{ashgriz1995temporal} where CFD Navier-Stokes simulations have been used for the same test-case.
In view of the results displayed in Fig.~\ref{fig:bu_time} and Fig.~\ref{HYDROCOMPARISON} one then asks where the mismatch between the two collision operators emerges, i.e. whether it is in the initial stage (where we are linearly unstable) or later at the pinch off stage. To answer this question we inspected the perturbation growth rate in the initial stage of the ligament destabilization and also considered the ligament \textit{silhouettes} for the whole dynamics up to break-up. Results are reported in Fig.~\ref{profiles}. We observe that the initial destabilization process is the same and well in line with the prediction of sharp interface hydrodynamics. The difference between the two dynamics rather lies in the pinching regime.
To have a deeper understanding of the ligament deformation near the break-up we also look at the non-dimensional "minimum" ligament radius as a function of the non-dimensional time. Results are reported in Fig.~\ref{fig:linear_pinch}. The "minimum" ligament radius is defined as the smallest radial coordinate in a configuration at a given time. We further make this quantity dimensionless by normalizing with the initial ligament radius. Notice that sharp interface hydrodynamics predicts a linear trend of the minimum ligament radius~\cite{eggers1993universal,lister1998} and the linear trend is more in line with the SRT dynamics than the MRT. It is worth nothing that the above presented method is characterized by a diffuse interface which tends to occupy few lattice points. Even though with increasing resolution the interface thickness tends not to influence results reliability, it is also true that for the maximum radius here considered (namely $R_0 = 98$ LU) the interface still occupies 4 LU. Here we consider that the interface is "physically" located exactly in the middle of the interface thickness. Thus, the LBM solver may be compared with the sharp interface hydrodynamics while the interface width $h\left(x,t^\star\right)$ is greater than the half of interface thickness (about 2 LU). Below that threshold there is no possibility to compare these two methods and comparison would be formally incorrect. It is important to point out, that the presented solvers do not introduce any special treatment so to reduce the interface thickness. Usually, such models are characterized by interfaces which occupy few sites, but some technicalities have been developed so to reduce the number of nodes occupied by the interface. Moreover, this thickness cannot be zero, like a sharp interface solver, and, its tuning will represent an additional degree of freedom while approaching these simulations. In the present study, we have decided to modify the ligament radius while keeping unchanged the interface thickness rather than modify the lattice sites occupied by the interface with the same ligament radius.
To delve deeper into the problem, we compare the velocity profiles for the ligament during the time evolution of the surface instability leading to the liquid column break-up. Results are displayed in Fig.~\ref{vel}. The Figure displays the dimensionless velocity magnitude $v* = \frac{|\textbf{u}^{b}|}{R_0/t_{\mbox{\tiny cap}}}$ for both collision operators for selected times. To facilitate a comparison between velocity distribution and the curvature profile, we compare the spatial distribution of velocity field in both MRT and SRT at similar interface morphology. We observe that the dimensionless velocity profiles are different for MRT and SRT, both in spatial distribution and magnitude. This difference in velocity greatly pertains the vapor phase, and the velocity distributions are differently correlated to the curvature: while for SRT the velocity localizes in the region of maximum curvature, for MRT it localizes in the regions of maximum curvature changes. Moreover, when curvature increases, the velocity contributions increase. A quantitative assessment of how much the observed velocity distribution is "spurious" would require the solution of the full hydrodynamic equations in presence of a vapor phase~\cite{eggers1994drop}. Nevertheless, the difference between SRT and MRT emerging from Fig.~\ref{vel} suggests that both collision operators lead to spurious contributions that develop "dynamically" and whose spatial localization is different for the same curvature profiles. Specifically, if the spurious currents on MRT are more localized at the pinch-off region, this can cause the break-up time to be different and the satellite droplets to be smaller after break-up. One could then reconsider the comments on the hydrodynamic recovery via the Chapman-Enskog theory: while SRT has extra terms with respect to MRT in its Chapman-Enskog expansion, the mismatch may be originated by the fact that the impact of spurious currents is more \textit{effective} in the present MRT implementation. In other words, the bulk equations are more correct in MRT than in SRT, but the interface boundary conditions are "dynamically" more "spurious" in MRT than in SRT. One could think of readsorbing this effect of spurious current in a modified stress tensor so that the dynamics of MRT would be that of a system with a slightly different Ohnesorge number. At the largest resolution analyzed, however, this would not be possible: we do not observe any steady satellite droplets, while theory predicts them to exist~\cite{driessen2013stability}.
\textcolor{black}{Summarizing, for the specific test case analyzed}, the pinching of the interface generates momentum and it is significantly influenced by the distribution of spurious currents. As we may observe from Fig.~\ref{vel}, the MRT with this specific set of parameters shows a stronger concentration of velocities in the pinching region, while the SRT presents weaker interference. We think this is the reason why the break-up dynamics captured from the SRT better matches the expected results, despite presenting some possible lacks of consistency while reconstructing the NS equation.
\section{CONCLUSIONS}\label{sec:conclusions}
We have investigated the dynamics of a liquid ligament perturbed via a Plateau-Rayleigh instability by means of the LBM. More specifically, we have considered two LBM collision operators, SRT and MRT, implemented in a multiphase numerical scheme based on the Shan-Chen Equation of State (EoS)~\cite{shan1994simulation,:ShanChen,Falcucci}. We have seen that numerical simulations display a neat asymptotic behavior, in the limit where the interface thickness is sensibly smaller than the characteristic radius of the liquid ligament. Such a behavior has been compared with the predictions of sharp-interface hydrodynamics~\cite{eggers1993universal,lister1998}, for both LB environments. Adopting the same EoS, the two collision operators displayed a different behaviour, with the SRT granting results more adherent to the theoretical and experimental evidence from the literature, compared to MRT. In particular, even if the break-up dynamics presents a very similar trend between the two collision operators in the early stages of the instability, SRT provides a pinching evolution closer to the theoretical predictions, with the eventual formation of long-living secondary droplet after the ligament break-up, which we have not detected in the MRT environment. This difference is traced back to the "dynamic" distribution of spurious currents rising in the pinching region. For the specific realizations of EoS adopted, the MRT -- despite a higher numerical stability granted in the set of hydrodynamical parameters characterizing the simulation -- displays a spurious currents pattern localized towards the flex of the ligament \textit{silhouette}, in the pinching region; SRT, on the other hand, is characterized by a different distribution of the spurious velocities, which appear to be localized away from the pinching region, hence the dynamics of the ligament break-up is less affected by such a spurious pattern and provides results closer to the sharp-interface hydrodynamics.\\
On a more general perspective, some comments are in order. Our results show that whenever spurious currents effects are weak in the pinching region, the LBM results can quantitatively match the ones obtained through the sharp-interface hydrodynamics, from the initial perturbation destabilization up to the break-up point. The fact that our simulations with SRT perform better than MRT may well depend on the specificity of the parameters chosen here, causing "dynamical" spurious currents to distribute less in the pinching region. While the issue of spurious currents has been pretty well detailed for "static" problems \cite{shan2006analysis,wagner2006thermodynamic,yuan2006equations,sbragaglia2007generalized,huang2011forcing,siebert2014consistent}, very little is known about "dynamical" problems. Hence, we do not want to "promote" a collisional operator rather than the other; instead, we want to point out that different "dynamical" distributions of spurious currents may produce quantitatively different results. Of course, this is just a ``macroscopic'' property which obviously hides many non-trivial dependencies on the parameters and technical details of the models used. This also opens up future perspectives in determining the impact of the different parameters/choices at hand (e.g. EoS, thermodynamic consistency, surface tension coupling with EoS, collisional scheme, interface width, Knudsen effects, etc.) to obtain a ``unifying view'' on what are the causes behind the emergence of "dynamical" spurious currents.
\section{ACKNOWLEDGEMENTS}
The numerical simulations were performed on \textit{Zeus} HPC facility, at the University of Naples "Parthenope"; \textit{Zeus} HPC has been realized through the Italian Government Grant PAC01$\_$00119 "MITO - Informazioni Multimediali per Oggetti Territoriali", with Prof. E. Jannelli as the Scientific Responsible. This project has received funding from the Italian Government Program PRIN grant n. 20154EHYW9. MS and XX thank the European Union’s Horizon 2020 research and innovation programme under the Marie Sk\l{}odowska-Curie grant agreement No 642069 for support.

\nocite{*}
\bibliographystyle{aipnum-cp}
\bibliography{sample}

\end{document}